\documentclass[10pt,letterpaper,twocolumn]{article}
\usepackage{ol2}
\usepackage[draft]{hyperref}
\usepackage[latin9]{inputenc}
\usepackage{amsmath}
\usepackage{amssymb}
\usepackage{graphicx}

\makeatletter
\@ifundefined{textcolor}{}
{%
 \definecolor{BLACK}{gray}{0}
 \definecolor{WHITE}{gray}{1}
 \definecolor{RED}{rgb}{1,0,0}
 \definecolor{GREEN}{rgb}{0,1,0}
 \definecolor{BLUE}{rgb}{0,0,1}
 \definecolor{CYAN}{cmyk}{1,0,0,0}
 \definecolor{MAGENTA}{cmyk}{0,1,0,0}
 \definecolor{YELLOW}{cmyk}{0,0,1,0}
}

\usepackage{epsfig}\usepackage{color}%\usepackage{colortbl}
\def\be{\begin{equation}}
\def\ee{\end{equation}}
\def\bea{\begin{eqnarray}}
\def\eea{\end{eqnarray}}
\def\bes{\begin{subequations}}
\def\ees{\end{subequations}}

\newcommand{\PT}{\mathcal{PT}}

\begin{document}

\twocolumn[

\title{$\PT$-symmetric coupler with a coupling defect: soliton interaction with exceptional point}

\author{Yuli V. Bludov$^{1}$, Chao Hang$^{2}$, Guoxiang Huang$^{2}$, and
Vladimir V. Konotop$^{3}$}

%\affiliation{
\address{
$^{1}$Centro de F\'{i}sica, Universidade do Minho, Campus de Gualtar,
Braga 4710-057, Portugal
%} \affiliation{
\\
$^{2}$ State Key Laboratory of Precision Spectroscopy and Department
of Physics, East China Normal University, Shanghai 200062, China
%}\affiliation{
\\
$^{3}$Centro de F\'{i}sica Te\'orica e Computacional Faculdade de Ci\^ecias, Universidade de Lisboa, Instituto
para Investigação Interdisciplinar, Avenida Professor Gama Pinto 2,
Lisboa 1649-003, Portugal,  and Departamento
de F\'{i}sica, Faculdade de Ci\^ecias, Universidade de Lisboa,  Campo Grande, Ed. C8, Piso 6, Lisboa 1749-016,  Portugal}

\date{\today}

\begin{abstract}

We study interaction of a soliton in a parity-time ($\PT$) symmetric coupler
%consisting of two \textcolor{red}{channels},
which has local perturbation of the coupling constant. Such a
%coupling
defect does not change the $\PT$-symmetry of the system, but  locally can achieve the exceptional point. We found that the symmetric solitons after interaction with the defect either transform into breathers or blow up. The dynamics of anti-symmetric solitons is more complex, showing domains of successive broadening of the beam and of the beam splitting in two outwards propagating solitons, in addition to the single breather generation and blow up. All the effects are preserved when the coupling strength in the center of the defect deviates from the exceptional point. If the coupling is strong enough the only observable outcome of the soliton-defect interaction is the generation of the breather.

\end{abstract}

%\pacs{...}
]

%\maketitle

Two coupled waveguides (a coupler), with gain and losses which are mutually balanced is a parity-time ($\PT$)-symmetric system~\cite{Ruter}. In the nonlinear case~\cite{Ramezani}  they represent a testbed for various phenomena involving instabilities and optical solitons. Such couplers support stable propagation of bright~\cite{Driben1,Driben2,Barashenkov1} and dark~\cite{dark} solitons, breathers~\cite{Barashenkov2}, and rogue waves~\cite{rogue}. The dynamical properties of these systems are determined by the relation between the strengths of the coupling ($\kappa$) and the gain-loss coefficient ($\gamma$),  splitting the region of the parameters in two domains corresponding to the unbroken $\PT$-symmetric phase, when all linear modes propagate without amplification or attenuation, and the domain where the linear modes are unstable (the broken $\PT$-symmetry). The value of the relation $\gamma/\kappa$ separating these two domains is an exceptional point (for discussion of physical relevance of exceptional points see e.g.~\cite{Heiss}).

When
%the ratio between
the coupling and gain/loss coefficients changes along the propagation distance, the  properties of the medium are affected and new effect can be observed.
%, suggesting tools of manipulate soliton dynamics in $\PT$-symmetric %couplers.
In particular, $\PT$-symmetry with alternating sign can stabilize solitons~\cite{Driben2}; a $\PT$-symmetric defect with localized gain and loss results in switching solitons between the waveguides~\cite{AKOS}.  The "governing" ratio $\gamma/\kappa$ can also be changed by varying the coupling coefficient.
%Practically
This can be done by changing the properties of the medium between the waveguides or by using curved waveguides with varying distance between the waveguides. Such situation was considered for conservative couplers in~\cite{Malomed1,Malomed2%,Malomed3
}, where local change of the coupling constant does not affect qualitatively the properties of the system. In the case of a $\PT$-symmetric coupler, however,
%one encounters an interesting situation:
if change of $\kappa$ locally reaches (or crosses) the exceptional point the properties of the coupler are changed qualitatively. In this case the $\PT$-symmetric phase is broken locally and one can speak about {\em exceptional point defect}.

One can expect that if the exceptional point defect is long enough [compared to the wavelength of soliton in the longitudinal direction], a soliton incident on it should become unstable. Indeed, in the spatial domain of the defect, a soliton cannot exist. Then one may expect different scenarios of the soliton instability. These scenarios are addressed in the present Letter. More specifically , we study of the interaction of a vector soliton in a $\PT$-symmetric coupler with the localized defect of coupling and  report four possibilities of the soliton evolution interacted with the defect: the excitation of a one-period breather, excitation of a breather with oscillating amplitude and width, the splitting of a vector soliton in two breathers, and intensity blowup.
%(i.e. infinite growth with time).

We consider two coupled waveguides described by
two nonlinear Schr\"{o}dinger equations
%~\cite{RTW}
\begin{eqnarray}
\label{eq:coup-PT}
\begin{array}{l}
iq_{1,z}=-q_{1,xx}+i\gamma q_{1}-\kappa(z)q_{2}-|q_{1}|^{2}q_{1},\\
iq_{2,z}=-q_{2,xx}-i\gamma q_{2}-\kappa(z)q_{1}-|q_{2}|^{2}q_{2}.
\end{array}
\end{eqnarray}
with the coupling
$
\kappa=\kappa_{0}-\left(\kappa_{0}-\kappa_{min}\right)e^{-z^{2}/\ell^{2}},$
characterized by the amplitude $\kappa_{0}-\kappa_{min}$ (i.e. it attains the minimal value $\kappa_{min}$ at $z=0$ and
tends to $\kappa_{0}$ at $z\to\pm\infty$) and by the width $\ell$.
To reduce the number of parameters
%(without loss of generality)
we set $\gamma=1$ and leave as the only controlling parameters, the ones describing the coupling defect, i.e. $\kappa_0$, $\kappa_{min}$ and $\ell$. Respectively, $\kappa_{min}=1$ corresponds to the exceptional point defect.

In the limiting region where, $\kappa_{min}\approx\kappa_0,$
%or in the case of constant coupling,
Eqs.~(\ref{eq:coup-PT}) possess a soliton solution~\cite{Driben1}
\begin{equation}
q_{1}^{ (\sigma) }=\frac{\sqrt{2}\eta\exp\left[i(\eta^{2}+ \sigma\kappa_{0}\cos\delta)z\right]}{\cosh\left(\eta x\right)}=\sigma q_{2}^{(\sigma)}e^{-i\sigma \delta},
\label{eq:solution}
\end{equation}
where $\delta=\arcsin\left(\gamma/\kappa_{0}\right)$ such that $0\le\delta\le\pi/2$. The soliton is parametrized by the positive parameter $\eta$,
and represent symmetric ($\sigma =1$) and  antisymmetric ($\sigma =-1$) solutions. Eq. (\ref{eq:solution}) at $z=z_{init}$ is used below for the initial data for vector solitons interacting with the defect.

\begin{figure}
\includegraphics[width=\columnwidth]{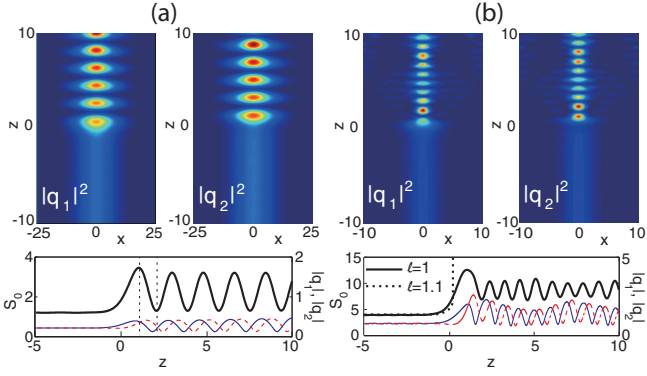}
\caption{(color online) Upper panels: Field intensities
%$|q_{1,2}|^2$,
with (a) $\eta=0.15$ and (b) $\eta=0.5$ interacting
with defect  at $z=0$. The coupling  $\kappa_{min}=1$,
$\kappa_0=2$ (a) and $4$ (b). Lower panels: Respective evolution
of the total energy flow
$S_0$ for $\ell=1$
[thick solid lines] and  soliton amplitudes $|q_1|$ and $|q_2|$ [thin solid
and dashed lines, respectively]. Thick dotted line in (b) shows
blowup at $\ell=1.1$.  The local maxima (minima) of $S_0$ [the vertical  lines in lower panel (a)]
happen where  the powers in the waveguides are
equal:
$\int |q_1|^2dx=\int |q_2|^2dx$.  The simulations for bounded
solutions have been performed between
$z_{ini}=-10$  and  $z_{fin}=100$ and
on the grid $-40<x<40$. \label{fig1}}
\end{figure}

Starting with the interaction of a symmetric soliton ($\sigma =1$)
with the exceptional point defect, $\kappa_{min}=1$, in
Fig.~\ref{fig1} we resume the typical results. The figure reveals
the two different dynamical scenarios, which depend on whether the
length of the defect $\ell$ is below or above some critical value
$\ell_{cr}$. In Fig.~\ref{fig1} (a) the soliton passes through a
relatively short defect transforming into a breather. The defect
width in this case, $\ell=1$,  is far below the critical value:
for $\eta=0.15$, $\kappa_0=2$, and $\kappa_{min}=1$ we found
$\ell_{cr}\approx 7$.
The emergent breather solution is characterized by the intensity
oscillations between the two components -- minimum (maximum) in
one component corresponds to maximum (minimum) in the other one
[Fig.~\ref{fig1} (a)].  The frequency of these oscillations (after
soliton passed the defect) can be estimated as
$2\sqrt{\kappa_0^2-\gamma^2}$.  For the weakly nonlinear limit
this estimate was derived in~\cite{Barashenkov2} (it stems from
the difference of the eigenfrequencies of the linear
$\PT$-symmetric coupler). At a finite amplitude the estimate for
the frequency can be obtained from the following arguments.
Introducing the Stokes variables $s_0=|q_1|^2+|q_2|^2$, $s_1=q_1
q_2^*+q_1^*q_2$, $s_2=-i(q_1 q_2^*-q_1^*q_2)$, and
$s_3=|q_1|^2-|q_2|^2$, as well as their integrals
$S_j=\int_{-\infty}^{\infty}s_j(z,x)dx$, we obtain
\begin{eqnarray*}
\label{Stokes}
\begin{array}{c}
\displaystyle{\frac{dS_0}{dz}=2\gamma S_3},\,\,\,
\displaystyle{\frac{dS_2}{dz}=-2\kappa (z) S_3+\int_{-\infty}^{\infty}\!\!\! s_1s_3dx},
\\
\displaystyle{\frac{dS_1}{dz}=-\int_{-\infty}^{\infty}\!\!\! s_2s_3dx},
\,\,\,
\displaystyle{\frac{dS_3}{dz}=2\gamma S_0 +2\kappa (z) S_2 }
\end{array}
\end{eqnarray*}
%(which however is not closed with respect to the integral Stokes components).
%Obviously, $S_0$ represents the total energy flow.
For $\eta\ll 1$ we have $\int |q_j|^4dx\sim\eta^2\int |q_j|^2dx$ and
$ \left|\int s_1s_3dx\right|=\left|\int |q_1|^4dx-\int |q_2|^4dx\right|\ll \left|S_3\right|. $
In the case at hand $\eta=0.15$ and $\kappa_0=2$ and the
nonlinear term in the equation for $S_2$ can be neglected with the
accuracy $\eta^2/\kappa_0\approx 0.011$. As a result the system   for $S_0$, $S_2$ and $S_3$ become closed and linear.
One of its eigenfrequencies is
$2\sqrt{\kappa_0^2-\gamma^2}$ giving period of
oscillations
$\pi/\sqrt{\kappa_0^2-\gamma^2}\approx1.8$; it agrees well with
the numerical results in Fig.~\ref{fig1} (a).

In Fig.~\ref{fig1} (b) the solution passes through the same defect
($\ell=1$) just below the critical value (for $\eta=0.5$,
$\kappa_0=4$, and $\kappa_{min}=1$ we found $\ell_{cr}\approx
1.1$) and is transformed into a breather. Now the period of
oscillations is $\pi/\sqrt{\kappa_0^2-\gamma^2}\approx0.8$, which
still agrees well with the numerical results. The dependencies of
the total energy flow $S_0$  and the solution amplitudes
$|q_{1,2}|$ on $z$ for each case are shown in the lower panels.   %\textcolor{red}
{When the defect width is close to the threshold value [Fig.\ref{fig1}], dependence $S_0(z)$
%looses its periodicity and
becomes  quasiperiodic.}

\begin{figure}
\includegraphics[width=\columnwidth]{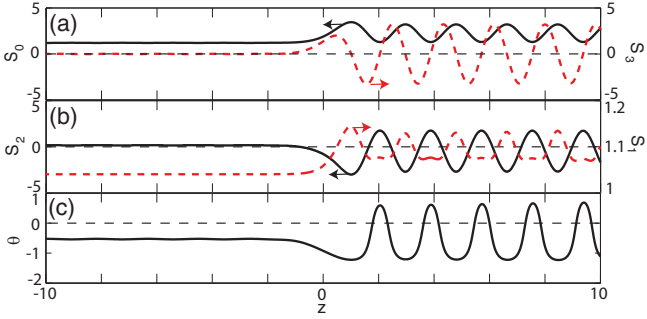}
\caption{(color online) (a): $S_0$ (solid line) and $S_3$ (dashed line)
{\it vs} $z$; (b): $S_2$ (solid line) and $S_1$ (dashed
line) {\it vs} $z$; (c):  $\theta$ {\it vs} $z$. The parameters are the same with those used in Fig.~\ref{fig1} (a). \label{fig2}}
\end{figure}

In Fig.~\ref{fig2} we show details of the evolution of the
Stokes components and phases of the emergent breathers.  The
breathing character of the mode is evident from almost periodic
power imbalance $S_3$ between the waveguides. We also
observe that the breathing solution is accompanied by the
oscillation of the "current" $S_2$ (which is constant for the
soliton solution). These oscillations are related to the lifting
the phase locking between the components [Fig.~\ref{fig2}]: the
phase difference $\theta=\arg q_1-\arg q_2$, which is constant for
 soliton (\ref{eq:solution}), in the breather solution depends
periodically on the evolution coordinate. We also confirmed that
the Stokes component $S_1$ remains mach smaller than the other
ones, what corroborates with the suppositions made in the
estimates of the breather period.

\begin{figure}
\includegraphics[width=\columnwidth]{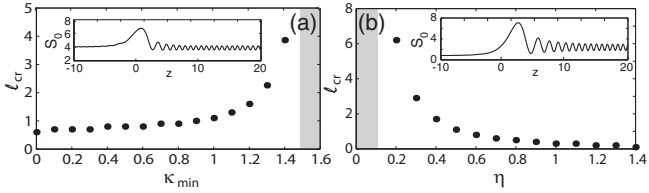}
\caption{The dependencies of $\ell_{cr}$ {\it vs} $\kappa_{min}$
for $\eta=0.5$ (a) and {\it vs} $\eta$ for $\kappa_{min}=1$ (b).
In both panels $\kappa_0=4$. If
$\kappa_{min}>\kappa_{min}^*\approx 1.5$ (a) and
$\eta<\eta^*\approx 0.1$ (b) (the gray domains) no blow-up is
found under the given values of parameters. Insets show the
dynamics of Stokes components for a soliton interacting with a
strong coupling defect $\kappa_{min}=1.5$ (a) and  for a small
amplitude soliton ($\eta=0.1$) interacting with the exceptional
point defect (b), where sufficiently long defect, $\ell=10$,
results in excitation of a breather. \label{fig3}}
\end{figure}

If the length of the defect exceeds a critical value $\ell_{cr}$
for a given coupling constant, the soliton "cannot overcome" it:
the component with gain $q_1$ (and hence $S_0$) grows infinitely. Thus the soliton after passing through the
defect blows up [see the dotted line in the lower panel of
Fig.~\ref{fig1} (b)]. We performed detail study of the dependence
of the critical width of the defect $\ell_{cr}$ as a function of
the minimal coupling $\kappa_{min}$ [Fig.~\ref{fig3} (a)]. The
main qualitative result is that the exceptional point
$\kappa_{min}=1$ separates quasi-linear (at $\kappa_{min}<1$) and
quasi-exponential (at $1<\kappa_{min}<{\kappa_{min}^*}$)
dependencies $\ell_{cr}(\kappa_{min})$.
Interestingly, when $\PT$-symmetry is locally broken
($\kappa_{min}<1$)  or even approaches zero, soliton still can
passe the coupling defect provided the defect is narrow enough.
At the same time, relatively strong coupling prevents blow up: for
$\kappa_{min}>{\kappa_{min}^*}$ there is no critical width of a
defect, and a soliton can pass a defect of \textit{any} width
being transformed in a breather. In the inset of Fig.~\ref{fig3}
(a) we show an example of strong coupling $\kappa_{min}=1.5$,
where the defect with sufficiently long width $\ell=10$ results in excitation of breathers. The blow up can occur in
the whole interval of weak coupling $0<\kappa_{min}<
\kappa_{min}^*$ (in Fig.~\ref{fig3} (a), $
\kappa_{min}^*\approx 1.5$ ).

In Fig.~\ref{fig3} (b) we show the dependence of $\ell_{cr}$ on
the inverse soliton width $\eta$ at $\kappa_0=4$ and
$\kappa_{min}=1$. For a given defect width there exist a critical
soliton amplitude separating small amplitude solitons which pass
the impurity being transformed in breathers and large amplitude
solitons which blow up. We also observe an upper critical
amplitude  $\eta_{cr}^2=2\sqrt{\kappa_0^2-1}/3\approx 1.6$, above
which a soliton blows up independently of the width of the defect.
This last effect is a manifestation of the instability of large
amplitude solitons in a $\PT$-symmetric coupler~\cite{Driben1}.
Like in the previous case, solitons with $\eta<\eta^*\approx 0.1$
are able to pass the defect of any width without blow-up. In the
inset of Fig.~\ref{fig2} (b) we show an example of excitation of a
breather by a small amplitude solitons.
%incident on a  sufficiently long defect width.

\begin{figure}
\includegraphics[width=\columnwidth]{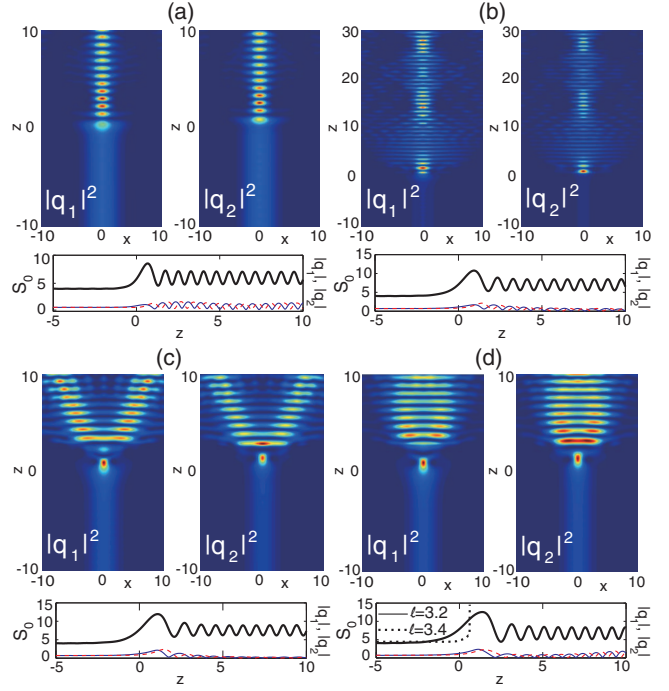}
\caption{(color online) Upper panels: The dynamics of soliton-defect interactions
for $\eta=0.5$ and $\ell=1.1$ (a), 2.2 (b), 2.7 (c), and 3.2 (d),
respectively, for the coupling $\kappa_0=4$ and $\kappa_{min}=1$.
In (a) and (b) The broadening is repeated along the propagation
distance with the period $\approx10$. Lower panels: The total energy flow $S_0$
[thick solid lines] and soliton amplitudes $|q_1|$ and $|q_2|$ [thin solid
and dashed lines, respectively] for each solution. Thick dotted line in (d) corresponds to the blow up happening at
$\ell=3.4$. \label{fig4}}
\end{figure}

Turning to the interaction of the antisymmetric soliton
$\sigma=-1$ with an exceptional point defect we observe  more rich
behavior, which is resumed in Fig.~\ref{fig4}.  As in the case of
symmetric soliton we find that there exists a critical defect
length $\ell_{cr}$ above which the soliton blows up (for the
chosen parameters $\ell_{cr}\approx 3.4$). If the width of the
defect is below $\ell_{cr}$, the soliton-defect interaction
results in creation of breathers, although this occurs now
according to different scenarios. The effect of a relatively short
defect acts  similarly on the symmetric and anti-symmetric
solitons, c.f. panels (a) in Figs.~\ref{fig1} and~\ref{fig4}, here
one observes that the antisymmetric breathers have
shorter period ($\approx0.8$) than that of the symmetric ones.

Increase of the defect lengths results in broadening of the soliton passed the defect [Fig.~\ref{fig4}
(b)]. This broadening is repeated along the propagation distance
[in Fig.~\ref{fig4} (b) the period $\approx 10$]. Further increase of $\ell$ leads to splitting of the incident soliton in the two outward propagating
pulses, as it is shown in Fig.~\ref{fig4} (c). It turns out that
the domain of the defect lengths leading to the splitting of the
incident beam is finite (for the parameters of Fig.~\ref{fig4}
this is the domain $2.2\leq\ell\leq 3.2$). Interestingly, further
increasing of the defect length stops soliton splitting and
reintroduces the scenario when broadening of the soliton is
observed [Fig.~\ref{fig4} (d)]. In spite of the reported diversity
of the behaviors, the total energy flow $S_0$ is increasing
smoothly with the growth of $\ell$ displaying no reflection of the
broadening or splitting dynamics.
%\textcolor{red}{(due to the fact, that $S_0$ is the integral parameter)}.

\begin{figure}
\vspace{-0.5cm}
\includegraphics[width=\columnwidth]{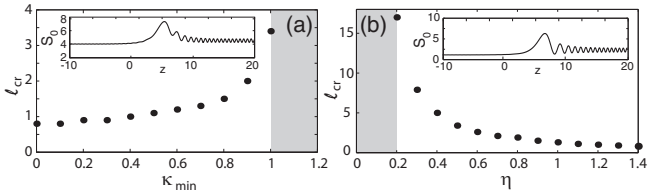}
\caption{(a) $\ell_{cr}$  {\it vs} $\kappa_{min}$
for $\eta=0.5$   and (b) $\ell_{cr}$ {\it vs} $\eta$ for $\kappa_{min}=1$.
In both panels $\kappa_0=4$. If $\kappa_{min}>
\kappa_{min}^*\approx 1$ (a) and $\eta< \eta^*\approx 0.2$ (b)
(the gray domains) no blowup occurs for the given parameters. Insets show the Stokes components for the
 defect with $\kappa_{min}=1.1$ (a) and for the small amplitude soliton ($\eta=0.1$) interacting with the exceptional point defect (b), where the defect of the length $\ell=10$  results in excitation of breathers. \label{fig5}}
\end{figure}

In Fig.~\ref{fig5} we show $\ell_{cr}$ {\it
vs} $\kappa_{min}$ [panel (a)]  and $\ell_{cr}$ {\it vs} $\eta$
for $\kappa_{min}=1$ [panel (b)] for the case of
anti-symmetric soliton. Comparing
Figs.~\ref{fig5} and~\ref{fig3} we observe that blowup of a symmetric soliton occurs at lower amplitudes and smaller defect lengths,
than the blowup of an anti-symmetric soliton.

Interactions of the solitons of both types with the defect obey
several common features. First, soliton-defect
interaction starts with the local increase of the energy flow.
Indeed, the initial (solitonic) values of the Stokes parameters are given by: $S_0^{(s)}=8\eta$, $S_1^{(s)}=8\eta\sigma\cos\delta$,
$S_2^{(s)}=-8\eta\sin\delta$, $S_3^{(s)}=0 $ ($s_3\equiv 0$) and
thus (\ref{Stokes}) gives that at the initial stage of evolution $S_0$ and $S_3$ are growing independently of defect parameters. Second, it follows from (\ref{Stokes}) that  for an {\em exact} breathing, i.e. $L$-periodic, solution   $\langle S_3 \rangle=\frac 1L \int_z^{z+L}
S_3(z)dz=0$. For a
breather far from the defect, where
$\kappa(z)\approx\kappa_0$, we also find that $\langle
S_2\rangle=-(\gamma/\kappa) \langle S_0\rangle<0$.  Thus, the
defect results in oscillations of $S_2(z)$ without changing
the sign of its average.

\begin{figure}
\vspace{-0.5cm}
\includegraphics[width=8.5cm]{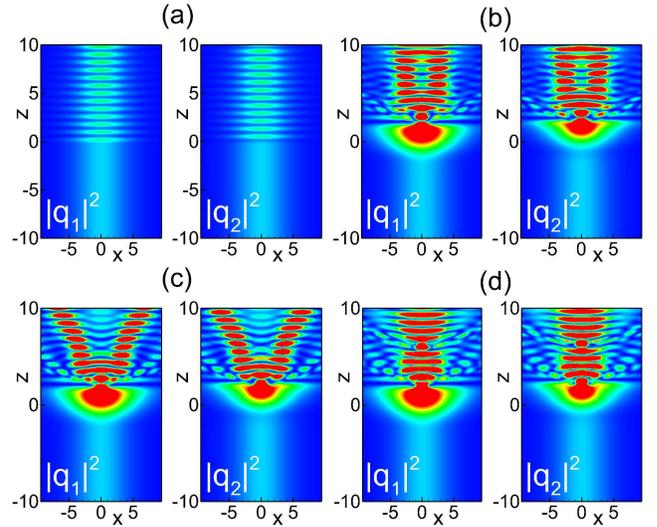}
\caption{(color online) Soliton interaction with super-Gaussian defect for $\eta=0.25$, $\kappa_0=4$, $\kappa_{min}=1$ and $\ell=0.2$ (a); $\ell=2.0$ (b); $\ell=2.1$
(c); $\ell=2.2$ (d).}
\label{fig:supergauss}
\end{figure}

Finally, using the super-Gaussian
defect $\kappa=\kappa_{0}-\left(\kappa_{0}-\kappa_{min}\right)e^{-z^{6}/\ell^{6}}$,
we checked how sensitive are our results to the choice of the defect. We found that for the parameters as in Fig.~\ref{fig1}(b) the critical value becomes $\ell_{cr}\approx0.5$. For
the antisymmetric mode results are shown in Fig.~\ref{fig:supergauss}.
We do observe that there are the same scenarios,
as those in Fig.~\ref{fig4} (although now $\ell_{cr}\approx2.3$ for $\eta=0.25$).
It is interesting, that for $\eta=0.5$ the critical value $\ell_{cr}\approx1.1$, i.e. considerably lower, than the one established in Fig.~\ref{fig4}.

To conclude, we considered interaction of a diffractive
soliton in a $\PT$-symmetric coupler with a coupling defect, which
locally achieves the exceptional point of the underline linear
system. Independently on whether the incident beam (soliton) is
symmetric or anti-symmetric, at relatively small defect length the
soliton passes through the defect and transforms into a breather.
This occurs even if in the region of the defect the $\PT$-symmetry
is broken. If the defect is long enough,  the total energy flow grows exponentially along the waveguides. In the case of an anti-symmetric soliton interacting with a defect there can exist domains where
successive broadening of the beam and even beam splitting
in two outwards propagating breathers occurs.

The work was supported by the Program of Introducing Talents of
Discipline to Universities under Grant No. B12024. YVB and VVK were
supported by FCT (Portugal) grants PEst-C/FIS/UI0607/2013, PEst-OE/FIS/UI0618/2011, PTDC/FIS-OPT/1918/2012.
CH and GXH were supported by the NSF-China
%through
grants 11105052 and 11174080.

\end{document}